\documentclass[usenatbib]{mnras}
\usepackage{graphicx}
\usepackage{amsbsy}
\usepackage{amssymb}
\usepackage{color}

\newcommand{\rd}{}

\def\araa{\rm{ARA\&A}}		
\def\apj{\rm{ApJ}}			
		
\def\apjs{\rm{ApJS}}		
		
\def\apss{\rm{Ap\&SS}}		
\def\aap{\rm{A\&A}}		
\def\aapr{\rm{A\&A~Rev.}}		
		
\def\mnras{\rm{MNRAS}}
\def\nat{\rm{Nat}}		
\def\physrep{\rm{Phys. Rep.}}

\def\solphys{\rm{Solar~Phys.}}

\def\note #1]{{\bf #1]}}

\definecolor{dred}{rgb}{0.85,0,0}

\definecolor{dblue}{rgb}{0,0,0.85}

\makeatletter
\let\oldcr=\@tabularcr
\makeatother

\title{On the hydrostatic stratification of the solar tachocline}
\author[J. Christensen-Dalsgaard, D.O. Gough and E. Knudstrup]
        {J. Christensen-Dalsgaard$^{1}$\thanks{e-mail: jcd@phys.au.dk},
        D.O. Gough$^{2}$\thanks{e-mail: douglas@ast.cam.ac.uk},
        E. Knudstrup$^{1}$\thanks{e-mail: emil@phys.au.dk}\\
	$^{1}${\rd Stellar Astrophysics Centre and} Department of Physics and Astronomy, Aarhus University,
              DK 8000 Aarhus C, Denmark\\
         $^{2}$Institute of Astronomy and Department of Applied Mathematics
              and Theoretical Physics, \\
              University of Cambridge, Cambridge CB3 0HA, UK}

\begin{document}

\maketitle
\label{firstpage}

\begin{abstract}
{We present an attempt to reconcile the solar tachocline glitch, a thin 
layer immediately beneath the convection zone in which the seismically 
inferred sound speed in the Sun exceeds corresponding values in standard solar 
models, with a degree of partial material mixing which we presume to have  
resulted from a combination of convective overshoot, wave transport and 
tachocline circulation.  We first summarize the effects of either modifying in the models the opacity in the radiative interior or of incorporating either slow {\rd or} fast tachocline circulation.  Neither alone is successful.  We then 
consider, without physical justification, incomplete material redistribution immediately beneath the convection zone which is slow enough 
not to disturb radiative equilibrium.  It is modelled simply as a diffusion 
process.  We find that, in combination with an appropriate opacity modification,  it is possible to find a 
density-dependent diffusion coefficient that removes the glitch almost 
entirely, with a radiative envelope that is consistent with seismology.}
\end{abstract}


\noindent
\section{Introduction}

There is some ambiguity in helioseismological inferences about the spherically averaged stratification of the Sun in the vicinity of the base of the convection zone. Consequently there is an associated uncertainty in the location of the base of the convection zone itself. Problems in interpreting helioseismological inversions of global frequency data arise because we are unsure of the dynamical consequences of convective overshoot, and of the influence of the magnetic field and of chemical-element segregation by gravitational settling moderated by the large-scale meridional flow in the tachocline. Issues such as these were recognized by \citet{jcddogmjtconvzonedepth1991ApJ...378..413C} in their original attempt to locate the base of the convection zone, causing them to suggest that the real uncertainty in their estimate is substantially greater than the formal precision of their data analysis, by a factor 3 or even more.

Christensen-Dalsgaard et al. adopted two different procedures for determining the radius $r_{\rm{c}}$ of the base of the convection zone: an absolute method, which used an asymptotic formulation of the entire sound-speed stratification of the star and which did not rely on a solar model, and a differential method, which used an asymptotic description of only the relatively small difference between the Sun and a standard theoretical model. The latter is naturally more precise, but it depends in a not wholly obvious manner on the approximations adopted in constructing the reference solar model. The conclusion from the differential method was $r_{\rm{c}} = 0.713 \pm 0.003 R$, where $R$  is the (photospheric) radius of the Sun.  Subsequently, \citet{basuantia1997MNRAS.287..189B} carried out a broadly similar asymptotic differential analysis with more precise frequency data, obtaining  $r_{\rm{c}} = 0.713 \pm 0.001 R$. It is essential to understand that the quoted errors indicate only the precision of the analysis, and may have little bearing on the accuracy of the calibration. Unfortunately, the distinction between the two has not always been appreciated.

Gyroscopic pumping associated with rotational shear in an essentially radiatively stratified layer beneath the convection zone \citep{eastachycline1972NASSP,easjpztach1992A&A} exchanges material in that layer with the convection zone, thereby opposing the tendency of gravitational settling to establish a gradient of helium and  heavier elements. This process, which must inevitably be taking place, although possibly in tandem with other mixing processes, tends to reduce abundance gradients on a timescale almost certainly less than the local characteristic settling time \citep{dogmem1998Nature}. The outcome is to make the spatial variation of the helium abundance $Y$ rather smoother than in standard theoretical models, resulting in a positive anomaly (acoustical glitch) in the difference between the sound speeds in the Sun and the models.  Such a glitch is evident in differential inversions of helioseismic data against standard solar models, as is illustrated in Fig.~\ref{fig1}, and has been recognized also by \citet{antiachitreseimicsun1998A&A..339..239A} as being a signature of material mixing .

If essentially laminar tachocline circulation were the only homogenizing agent, it would be possible in principle to determine the thickness of the tachocline from the properties of the acoustical glitch. A calibration of the dimensions of the glitch is likely to be more precise than analyses of rotational splitting, for example, because the mean multiplet oscillation frequencies that are employed have been determined from observation more precisely than has rotational splitting.   However, the outcome is not necessarily more accurate, because it is conditional on the assumptions of the stellar modelling, both on the physical approximations adopted in constructing the standard reference solar model and on the structure adopted for the tachocline, now and throughout the earlier evolution of the Sun. 

A calibration of the glitch was attempted by \citet{elliottdogtachthickness1999ApJ...516..475E} and \citet{elliottdogsekiitachocline1998ESASP}, who adopted what one might regard as the simplest of assumptions: namely that the stratification of the Sun is
 spherically symmetrical, that material mixing is complete down to the base of the tachocline at a rate that does not disturb radiative equilibrium, essentially in conformity with the tachocline model of Gough and McIntyre (1998), sometimes 
referred to as a slow tachocline, and that the thickness $\Delta$  of the tachocline has 
not changed throughout the main-sequence lifetime of the Sun, together, of course, with the other standard assumptions of solar modelling \citep{jcdetal1996Sci}.  They 
found that $\Delta=0.020 R$, the formal uncertainty in that value, resulting from only the quoted uncertainties in the seismic frequency data, being about 5$\%$.

Elliott and Gough were encouraged by their finding that not only the height and width of the acoustic glitch but also its functional form could be reproduced within the formal errors by adjusting but a single parameter $\Delta$.  That provided some, yet unjustified, modicum of faith in their result. However, the glitch in the calibrated model was displaced upwards by $0.015 R$ from that in the reference model, the consequences of which we discuss in the next section.   With what now appears to be the false confidence in their partial success, they presumed that the calibration of $\Delta$ was reliable, and that the fully mixed region was deeper than earlier analyses had indicated.  

It was never intended to be believed that the simplistic tachocline modelling of Elliot and his colleagues, based on the minimalist analysis of Gough and McIntyre (1998), should be strictly trusted. Other processes are also operative, not least of which is convective overshoot, which smears the mathematically precise location of both the base of the convection zone and the tachopause, beneath which in the simplistic model there was supposed to be no fluid motion. It was accepted that temporally varying overshoot of descending convective plumes must occur, but it was assumed that its long-term effect was small compared with that of the large-scale tachocline flow. In contrast, others have argued either that overshoot \citep[e.g.][]{Berthomieu_etal_convective_penetration1992ASPC...26..158B} or rotationally driven shear turbulence \citep[e.g.][]{richardvauclaircharbonneldziembowskisolarmodels1996A&A...312.1000R,brunt-czahnstandardsolarmodels1999ApJ} provides the principal source of material redistribution, possibly in the manner of an effective diffusion, and that it dominates the tachocline flow in smoothing the abundance profiles. No such assumption appears to have succeeded in removing the acoustic anomaly entirely \citep[e.g.][]{BACZ_tachocline_mixing2002A&A...391..725B,brunzahn2006A&A}.

Although the difference between the sound speed in the Sun and that in standard solar models is, by many astrophysical standards, extremely small, it is very much larger than the standard errors in the determinations of the sound speed (in places by as much as 20 formal standard errors in the helioseismological inversions). Therefore the discrepancies are very significant, and demand explanation.
Those discrepancies can quite easily be separated into two components: one that varies on a lengthscale comparable with the size of the Sun, and is no doubt a consequence of gross, albeit small, errors in properties that influence the global structure, such as the primordial chemical composition, the treatment adopted for the opacity which controls the radiative transfer of energy, or the dominant nuclear cross-sections, and one that is local, and is produced by errors in the treatment of processes that influence the sound speed directly.
The tachocline anomaly is the more prominent member of the latter component, and is the subject of the investigation reported here.

It has long been appreciated that a spatially confined sound-speed anomaly is most likely to be a result of an error in the pressure-density relation adopted in the reference solar model, whose influence on the sound speed $c=(\gamma_1 p /\rho)^{1/2}$ 
is direct. Here $p$ is pressure, $\rho$ is density and $\gamma_1=(\partial\,{\rm ln}\, p/\partial\,{\rm ln}\rho)_{\rm ad}$ is the first adiabatic exponent. Such an error could arise from an error in the concentrations of the abundant elements, which influences the mean molecular mass $\mu$.  In contrast, an error in the opacity or a controlling nuclear reaction rate influences the stratification only via coefficients that multiply derivatives in the governing differential equations,  and therefore has more spatially extensive consequences \citep[e.g.][]{elliottdogtachthickness1999ApJ...516..475E}.  
Although there is well appreciated uncertainty in the physics of the equation of state relating $p$ to $\rho$, which might well be  of a magnitude comparable with the sound-speed anomaly considered here \citep[e.g.][]{dappenetal_eoscomparison1990SoPh..128...35D, JCDWDEoS1992A&ARv, BaturinWDDOGSVV2000MNRAS}, it is quite unlikely that it would be restricted to a narrow region in the $p$--$\rho$ relation, 
particularly the region that happens (by chance, from a thermodynamical point of view) to coincide with the base of the solar convection zone. Therefore the most likely culprit is the spatial variation of the relative abundances of the most abundant chemical elements, helium 
and hydrogen, for only they have a sufficiently strong influence on the equation of state. Our task is therefore to examine that variation, a variation which is determined by the balance of overshoot, shear turbulence, 
wave transport 
and baroclinic tachocline flow {\rd against diffusion and settling}. Here we 
do not address the dynamical matters directly, but instead we attempt to determine the seismological consequences of putative mixing processes, and compare them with 
observation, in the hope that our conclusions will provide useful constraints on dynamical studies to come.

The large-scale global structure of calibrated solar models is 
influenced directly by the opacity, which controls the radiative interior, and indirectly by the location of the base of the convection zone, 
which is formally determined by conditions locally.
The latter was realized early in the days of helioseismology from comparisons
between theoretical solar models \citep[e.g.,][]{JCDhelioseismology1988}; 
the former came prominently into attention with the announcement by
\citet{asplundetal2004A&A} that the abundance $Z$ of heavy elements, 
and therefore the opacity produced from them, is substantially lower
than had previously been believed (at least in the photosphere and the well mixed convection zone), 
thereby augmenting the discrepancy between the seismologically 
determined stratification of the Sun and that of 
standard solar models, such as Model S of \citet{jcdetal1996Sci}.
An extended review of these issues is provided by
\citet{basu_antia_2008PhR...457..217B}.
Attempts to compensate for Asplund's revised composition have not been 
successful within the realm of currently accepted physics
\citep{guzik_mussack_2010ApJ...713.1108G}.
However, given that the composition affects predominantly the opacity,
modifications to opacity with no {\it a priori} physical basis
can be adjusted to cancel the 
effects of the composition change to bring models back to Model S
\citep{JCDetalopacity2009, JCD-Houdek2010},
leaving the tachocline anomaly intact.

To relate our current investigation to earlier studies, 
we first investigate simple attempts to bring Model S into better
agreement with the seismically inferred sound-speed stratification.
We then consider the effect on models computed with the Asplund et al.
composition of adjusting the opacity in the radiative interior 
and subsequently the degree of material mixing in the
tachocline, both without regard to how such adjustments might be
accomplished physically, in an attempt simply to reproduce that stratification.

%

\section{The tachocline anomaly}

\begin{figure}
          \includegraphics[width=8.0cm]{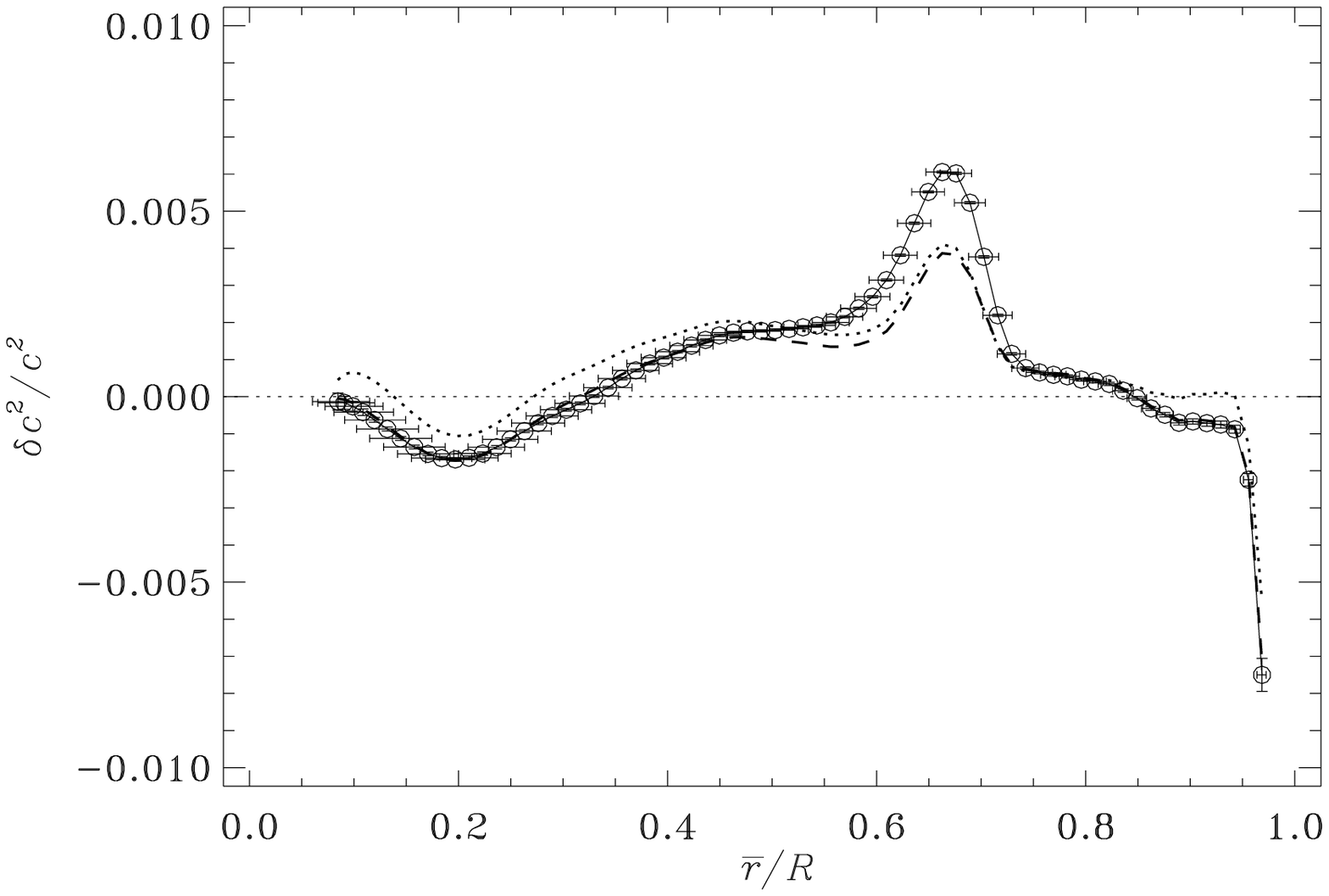}
\caption{The symbols depict optimally localized averages of the relative difference between the squared sound speed in the Sun and that in the new Model S,
in the sense (Sun) -- (model), which was constructed with the
opacities of \citet{updatedopacity1996ApJ.464.943I} and calibrated to $Z_{\rm s}/X_{\rm s} = 0.0245$, corresponding to the
heavy-element abundances of \citet{Grevessenoels1993};
they are plotted against 
the centres $\overline{r}$ of the averaging kernels, and joined by straight-line segments. The horizontal bars 
{\rd indicate} the widths of the averaging kernels, and the vertical bars denote standard errors arising from the uncertain errors, assumed to be independent, in the observational data \citep[obtained from][]{basu_etal_1997MNRAS.292..243B}.  The dashed curve is drawn through similar averages of the relative difference between $c^2$ in the Sun and that in the 
original Model S \citep{jcdetal1996Sci}, which was constructed with the
opacities {\rd of} \citet{opal92_rogers_iglesias, OPAL?1992ApJ...401..361R}
and calibrated also to $Z_{\rm s}/X_{\rm s} = 0.0245$.  
The dotted curve similarly shows results from an analysis of a model 
(Model S\'\,; see Section \ref{sec:seismod})
computed with the opacities of \citet{updatedopacity1996ApJ.464.943I}, 
the relative heavy-element abundances of \citet{Asplundetal2009ARA&A},
calibrated to $Z_{\rm s}/X_{\rm s} = 0.0181$ and including the
opacity modification of \citet{JCD-Houdek2010}.
}
\label{fig1}
\end{figure}

In Fig.~\ref{fig1} is depicted the difference in the squares of the sound speeds in the Sun and in a new standard solar Model S, the latter having been computed and calibrated in the same manner as the original Model S of \citet{jcdetal1996Sci}, 
including a present ratio $Z_{\rm s}/X_{\rm s} = 0.0245$
between the surface abundances by mass of heavy elements and hydrogen
{\rd \citep{Grevessenoels1993}},
but using the more modern OPAL opacities of  \citet{updatedopacity1996ApJ.464.943I}.%
\footnote{Regrettably, \citep{jcdetal1996Sci} did not make clear that the
\citet{opal92_rogers_iglesias, OPAL?1992ApJ...401..361R}
tables were used in the computation of Model S.}
The initial heavy-element abundance is $Z_0=0.0196$, the same as in the original Model S. For comparison we include a plot of differences from the original Model S.  Throughout this paper, the term Model S refers to the new model, unless explicitly stated otherwise. 
For completeness we note that the model is calibrated to
a photospheric radius of $6.9599 \times 10^{10}\,{\rm cm}$
and surface luminosity of $3.846 \times 10^{33} \, {\rm erg \, s^{-1}}$ 
at an age of 4.6\,Gyr.
The equation of state used the OPAL tables of 
\citet{rogers_etal_1996ApJ...456..902R} and nuclear parameters mostly from
\citet{Bahcall_etal_1995RvMP...67..781B}.
Diffusion and settling of helium and heavy elements, at the rate of
fully ionized oxygen, were included using the approximation of 
\citet{michaud_proffitt_1993ASPC...40..246M}.
Plotted in Fig.~\ref{fig1} are 
optimally localized averages \citep[e.g.][]{dog1985SoPh..100..65G,  dogmjtinversion1991_arizonabook, rabellosoaresetalinversionparameters1999MNRAS.309.35R}  of those differences, 
which are averages weighted with Gaussian-like weight functions (commonly called averaging kernels) centred about $\overline{r}$ with {\rd widths 
indicated by the horizontal bars;
specifically, these extend from the first to third quartiles 
of the averaging kernels,
with a separation of around 0.57 times the full width at half maximum of
the kernels.}
The vertical bars denote standard errors resulting from the quoted errors in the observed frequencies, computed under the assumption that the errors in the observations are independent.
The inverse analysis was carried out in terms of the pair $(c^2, \rho)$.
To facilitate comparison with earlier analyses
\citep[e.g.,][]{jcdmpdimdiffusion2007EAS..26..3C}, here and in the following 
we use the same observational data as did \citet{basu_etal_1997MNRAS.292..243B}.

We discuss first the prominent hump centred at $\overline{r}\simeq 0.67 R$  immediately beneath the convection zone where the tachocline is located.   
Elliott and Gough (1999) called it the tachocline anomaly, because they believed it to be a product of tachocline mixing, and we adopt that nomenclature here.

The discrepancy was postulated to have 
been caused by the tachocline circulation which returns settling helium to the convection zone, thereby locally augmenting the hydrogen abundance, and with it the sound speed. The fact that Elliott and Gough had found that the anomaly, although located too high in the envelope, could 
otherwise be reproduced by adjusting only the single parameter $\Delta$ led them  naively to suppose that the structure had been predicted robustly, and that 
the anomaly could simply be shifted downwards by global adjustments to the model that deepened the convection zone. Consequently they made no attempt to construct a 
fully self-consistent calibrated evolutionary model. However, as we demonstrate in the next section, a self-consistent model with a Gough-McIntyre tachocline that is not contradicted by seismology cannot be constructed in so simple a fashion. This should 
have been anticipated, because it has long been known that changing the depth of the 
convection zone in a model would introduce an additional sharp feature in the sound-speed difference 
\citep{JCDhelioseismology1988}, manifest as a discrepancy 
in the sound-speed gradient that is concentrated principally in and immediately beneath the region in which one model is radiative and the other is convective;  however, the sound-speed perturbation itself extends deep into the radiative interior. 
We discuss these aspects in some detail in the following sections.
To bring the model in line with seismic observation, 
alternative features, to which we return in Section \ref{sec:seismod},
need to be included in the model.

\section{The effect of slow tachocline mixing on hydrostatic stratification}

\begin{figure}
          \includegraphics[width=9.0cm]{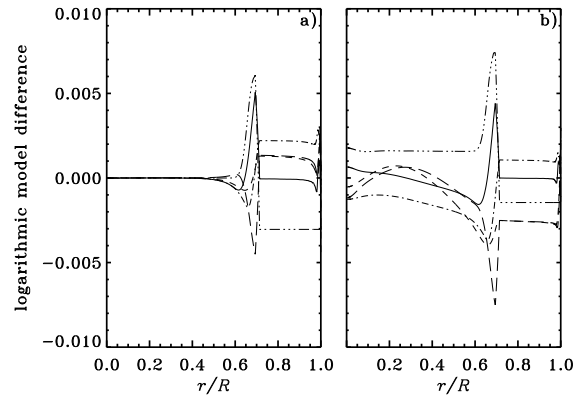}
\caption{Differences at fixed $r/R$ in ${\rm ln}\, c^2$ (continuous curve), ${\rm ln}\, p$ (short dashed), ${\rm ln} \,\rho$ (long dashed), ${\rm ln} \,T$ (dot-dashed) and $X$ (triple-dot-dashed) between the model with a slow tachocline, described in Section~3, and Model S.
The model in panel (a) was computed with the same $Z_0$ as Model S, 
namely {\rd 0.0196}; the model in panel (b) was calibrated to the same present value of $Z_{\rm s}/X_{\rm s}$,
namely 0.0245.
}\label{fig2}
\end{figure}

Fig.~\ref{fig2}b depicts {\rd deviations from Model S (which we denote by $\delta$ {\rd below}) at fixed $r/R$  of $\ln c^2$, $\ln p$, $\ln \rho$, $\ln T$,
where $T$ is temperature, and the hydrogen abundance $X$} in a properly calibrated solar model -- reproducing the observed values 
of the luminosity, ${\rm L}_\odot$, the radius, ${\rm R}_\odot$, and a heavy-element--to--hydrogen abundance ratio, $Z_{\rm s}/X_{\rm s}=0.0245$, in the photosphere -- having a slow tachocline modelled by 
mixing the chemical composition in a layer $0.02 R$ thick beneath and contiguous with the convection zone, yet retaining radiative equilibrium.
We note that, as detailed in Appendix A, the sound speed in
the bulk of the convection zone is largely fixed by the surface mass and radius
of the model, such that here $\delta c^2 \approx 0$.
Thus the mixing of otherwise gravitationally settled material,
causing a reduction in the hydrogen abundance in the convection zone with a consequent augmentation of the mean molecular mass $\mu$,
is compensated by a corresponding augmentation of $T$.
For comparison, we include a similarly constructed model with the same initial heavy-element abundance, 
$Z_0=0.0196$, as Model S; in that model the value of $Z_{\rm s}/X_{\rm s}$ is 0.0249, which differs from 0.0245 by much less than the observational uncertainty 
\citep{Grevessenoels1993}.

Had the initial heavy-element abundance $Z_0$ been held constant in the calibration, these modifications to the outer layers of the star would have had little influence on conditions deep in the radiative envelope and the energy-generating core, and all 
the perturbations would have declined towards zero (Fig.~\ref{fig2}a).  It is evident from the equation of 
hydrostatic support in the form ${\rm d\, ln}\,p/{\rm d}\,r=-\gamma_1 G m/c^2r^2$, 
where $m$ is the mass within a radius $r$ and $G$ is the gravitational constant,   that the raised sound speed in the acoustic glitch would have reduced the magnitude of the pressure gradient, thereby causing the pressure, and concomitantly the density, in the 
convection zone to be increased.   In the tachocline itself, $p/\rho$ would have been  increased, requiring an even greater inwards  reduction in ${\rm ln}\,\rho$ than in ${\rm ln}\,p$.  These changes induce lesser, compensating, deviations with reversed sign 
in a transition zone beneath, as depicted in Fig.~\ref{fig2}.

The reduced surface abundance $X_{\rm s}$ caused by the tachocline circulation leads to  a reduced heavy-element abundance $Z_{\rm s}$ in a calibration at fixed $Z_{\rm s}/X_{\rm s}$, a reduction  
which is transmitted throughout the entire star.   The resulting diminution of the opacity in the radiative interior leads to a global reduction in temperature, and a reduction of $p$ and $\rho$ in the outer radiative envelope and convection zone, enough to 
reverse the sign of $\delta {\rm ln}\,p$ and $\delta {\rm ln}\,\rho$ in the convection zone.    Consistent with the maintenance of hydrostatic support under homologous change, which approximately characterizes the modifications  undergone by  the recalibration, the reduction in $T$ is balanced principally by a corresponding reduction in $\mu$, 
therefore an augmentation in $X$, which maintains  
the nuclear reaction rates; the jump in $X$ across the tachocline is hardly changed, so there is some mitigation of the {\rd reduction in $X_{\rm s}$ by the recalibration}.

\begin{figure}
          \includegraphics[width=8.0cm]{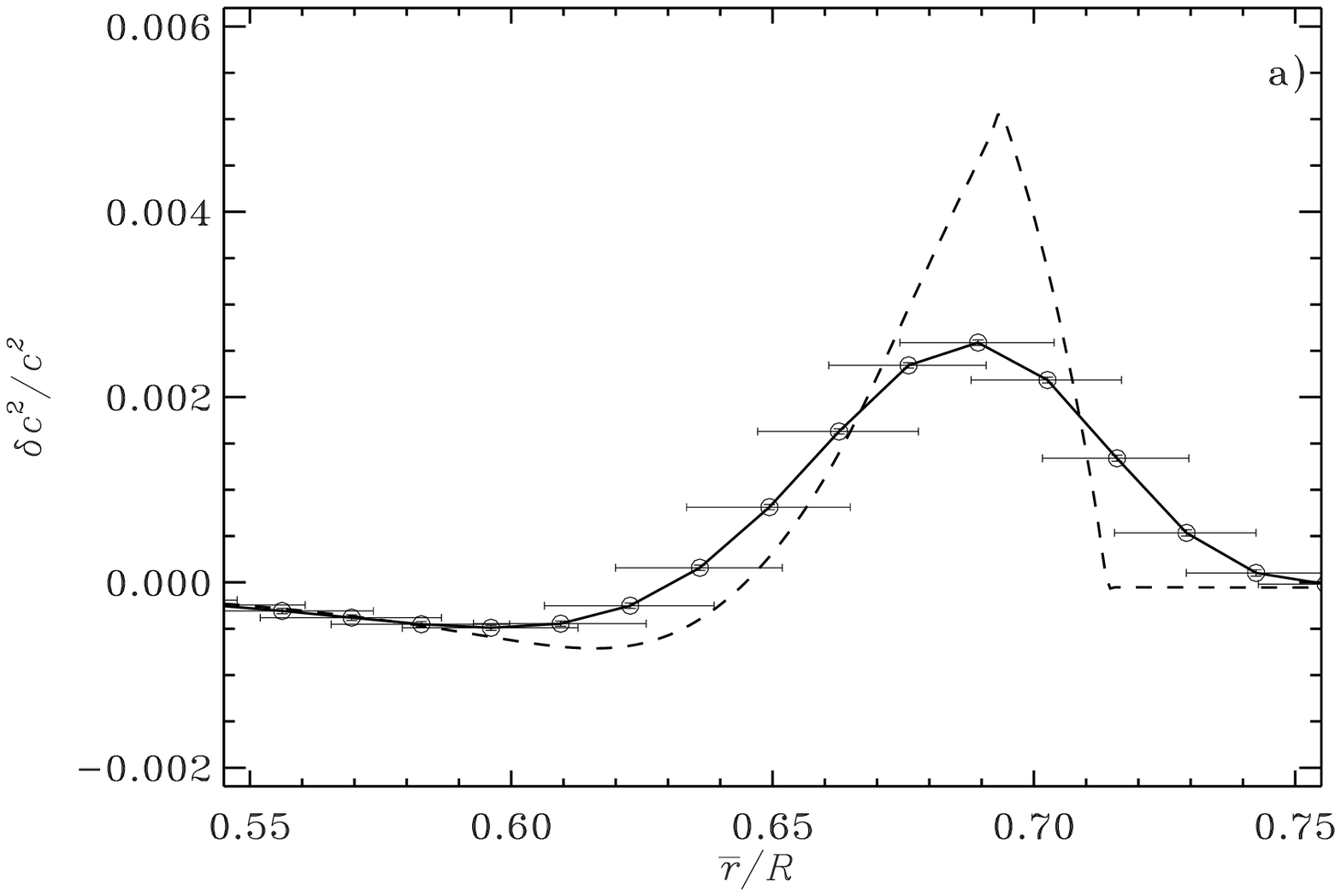}
          \includegraphics[width=8.0cm]{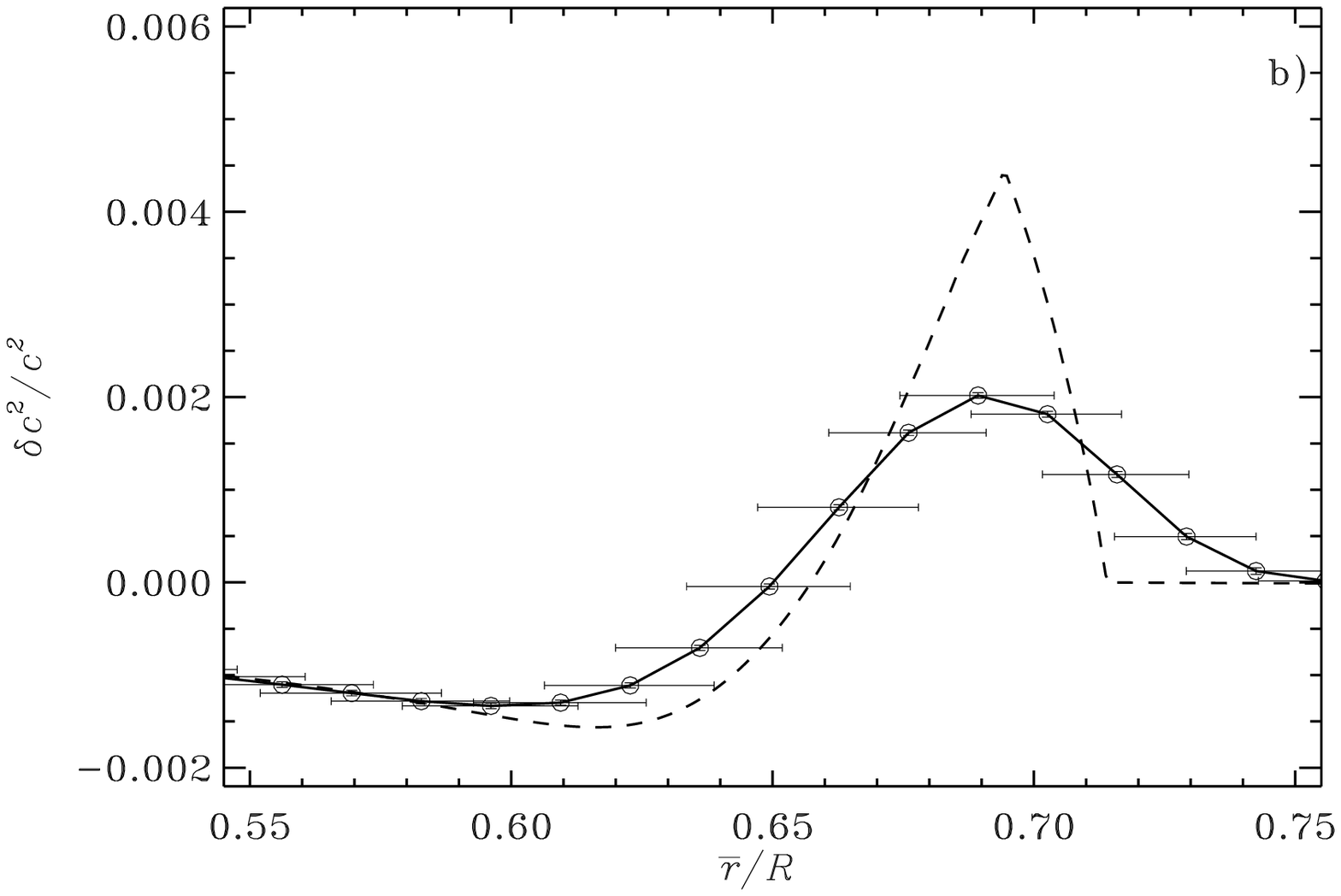}
	  \caption{The symbols joined by continuous line segments represent the squared sound-speed difference plotted {\rd in Fig.~\ref{fig2},
convolved with the optimal averaging kernels, and plotted against the centres of those kernels;
the original squared sound-speed differences are reproduced here as dashed curves.}
The model in panel (a) was computed with the same $Z_0$ as Model S, the model in panel (b) was calibrated to the same present value of $Z_{\rm s}/X_{\rm s}$.
}\label{fig3}
\end{figure}

When $\delta c^2$ is convolved with the averaging kernels of the inversion (Fig.~\ref{fig3}), the anomaly in this model appears to be centred higher in the envelope than the anomaly observed (see Fig.~\ref{fig1}), by a little over $0.015R$, at $r=0.685 R$,  
more-or-less as Elliott and Gough had found. 
{\rd This is because the anomaly is asymmetric with respect to $r$, 
and the kernel averages appear to broaden the steeper (upper) side of the
anomaly more than they do the lower.}
A direct consequence of that property is 
a small, yet sharp, dip in the sound-speed difference immediately beneath the 
convection zone of the Sun.  By what means can the location of 
{\rd the model} anomaly be lowered?

\section{Simple adjustments to the model tachocline}

One cannot simply imagine the adiabatically stratified region of the convection zone to be deepened without addressing how that could come about. Here we restrict ourselves to the class of proposals most commonly encountered in discussions of the internal 
structure of the Sun:  maybe there is non-locally driven,
{\rd almost adiabatic, motion}
penetrating more deeply than the level of neutral stability, in the 
form of `fast' large-scale tachocline flow or relatively local overshooting plumes that subsequently mix with their immediate surroundings; 
alternatively, for some reason, the opacity in the outer layers of the radiative zone is greater than is normally supposed, causing the local condition for 
convective instability to occur at a greater depth.  We consider the two  possibilities separately.  

\begin{figure}
          \includegraphics[width=9.0cm]{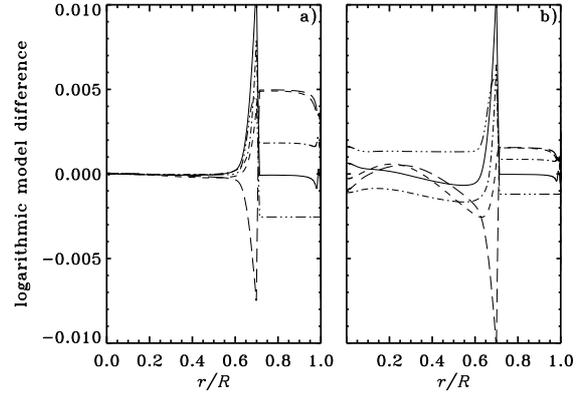}
\caption{Differences in ${\rm ln}\, c^2$ (continuous curve), ${\rm ln}\, p$ (short dashed), ${\rm ln} \,\rho$ (long dashed), ${\rm ln} \,T$ (dot-dashed) and $X$ (triple-dot-dashed) between the model with a fast tachocline, or
{\rd nearly adiabatic} overshoot,  and Model S.  The model in panel (a) was computed with the same $Z_0$ as Model S, the model in panel (b) was calibrated to the same present value of $Z_{\rm s}/X_{\rm s}$.
}\label{fig4}
\end{figure}

In Fig.~\ref{fig4} are plotted perturbations resulting from 
{\rd nearly adiabatic} overshooting (or a fast tachocline), modelled by artificially extending the adiabatically stratified region of the convection zone by $0.015 R$ 
and mixing 
the chemical composition with the convection zone.   Because the magnitude of the temperature gradient in the extended layer is greater than its radiative counterpart, the 
temperature itself is greater.  Therefore the sound-speed anomaly is greater than that in a model with a slow tachocline of the same depth. The greater sound speed in the 
acoustic glitch requires a pressure-gradient perturbation of greater magnitude, and a consequent greater increase of pressure and density in the convection zone, 
rendering the deviations positive in the convection zone even when $Z_{\rm s}/X_{\rm s}$ is held fixed in the recalibration.  The strong concentration of the acoustic anomaly 
near the base of the convection zone renders this simple well mixed model incompatible with seismic inference.

\begin{figure}
          \includegraphics[width=9.0cm]{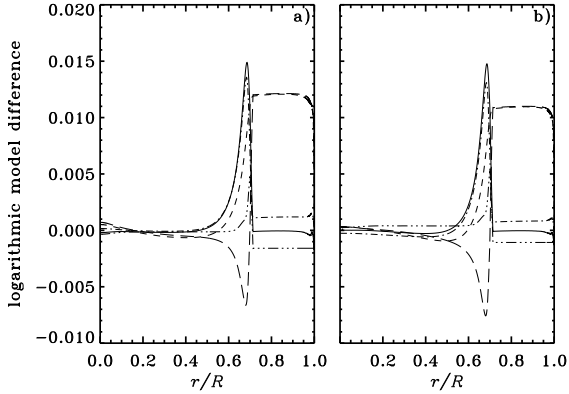}
\caption{Differences in ${\rm ln}\, c^2$ (continuous curve), ${\rm ln}\, p$ (short dashed), ${\rm ln} \,\rho$ (long dashed), ${\rm ln} \,T$ (dot-dashed) and $X$ (triple-dot-dashed) between the model in which the opacity has been artificially augmented by the factor $f$ 
{\rd (see equation \ref{eq:opmod})}
and Model S.  The model in panel (a) was computed with the same $Z_0$ as Model S, the model in panel (b) was calibrated to the same present value of $Z_{\rm s}/X_{\rm s}$.
}\label{fig5}
\end{figure}

In Fig.~\ref{fig5} are plotted perturbations resulting from artificially augmenting the opacity  by a factor $f$ in a thin layer beneath the convection zone, but not mixing the 
chemical species where the stratification is convectively stable  
\citep[cf.][]{tripahyjcdI_1998A&A...337..579T}.  
Specifically, the function $f$ is a 
constant, $f_0$, from where ${\rm log}\,T=6.4$, namely at $r = r_\kappa \simeq 0.68R$, up to the base of the convection zone, and declines beneath $r_\kappa$ 
{\rd on a characteristic lengthscale $0.005R$ according to:
\begin{equation}
f = \left\{ 
	\begin{array}{ll}
	f_0 & \mbox{for $r_\kappa < r < r_{\rm c}$} \\
	f_0 {\rm exp}[-2500({\rm log}\,T-6.4)^2] & \mbox{for $r < r_\kappa$} \; ;
	\end{array}
\right.
\label{eq:opmod}
\end{equation}
$f_0$ was taken to be 0.236,} that value having been chosen to make the 
convection zone $0.015 R$ deeper.    As should be expected, both $\delta\,{\rm ln} c^2$ and $\delta\,{\rm ln}T$ rise abruptly, and almost linearly, from zero at the base $r_{\rm{cS}}$ of the convection zone of Model S to the base $r_{\rm cc}$ of the convection 
zone of the corrupted model, because in the corrupted model the (now unstable, yet 
essentially) adiabatic stratification has been extended into the region where the magnitude of the temperature gradient in Model S is smaller.  However, the 
magnitude of the perturbation is too great.  In 
the extended region of the convection zone the chemical species 
are, of course, fully mixed in with the rest of the convection zone of the corrupted model, which 
accounts for the thin spike in the profile of $\delta X$.  Because the perturbation 
is localized in a region where the density is low compared with the mean, the structure deep in the star is hardly affected if $Z_0$ is held fixed. Then all the deviations  decline  
inwards below $r_{\rm cc}$ almost to zero, in common with the models illustrated in Fig.~\ref{fig2} and \ref{fig4}. If $Z_{\rm s}/X_{\rm s}$ is held fixed in the calibration, then the concomitant reduction in $Z$ in the radiative interior leads to a 
reduction in $T$, requiring an augmentation in $X$, similar to perturbations illustrated 
in Fig.~\ref{fig2} associated with the slow tachocline, although with a substantially lesser magnitude.  It should be recognized that the somewhat different shape to the 
acoustic anomaly that is produced by this device implies that the depth of 
the unstably stratified region is not necessarily extended by the apparent displacement of the slow-tachocline anomaly.  Therefore one might consider instead a different 
combination of magnitude and location of the opacity modification to produce an 
anomaly of the correct height.    However, the absence of the negative deviation 
$\delta c^2$ beneath the tachocline that is evident in Fig.~\ref{fig2} inhibits attempts to establish an overall 
fit of the model to the seismic data.  Moreover, the best fitting model requires an abrupt change in the opacity to be very close to the base of the convection zone, whose location has no straightforward connection with the atomic physics determining the  opacity.

\section{A search for a seismically acceptable model}
\label{sec:seismod}

The devices discussed in the previous section to deepen the convection zone introduce changes in the  sound speed that are concentrated near the base of the convection zone.  They have broadly similar  
characteristics very close to the convection zone, in that they both produce a positive 
acoustic anomaly.  The anomalies differ in their thickness--amplitude ratios, and  their extensions deeper into the radiative interior are different.  We 
are therefore moved to enquire whether an artificial modification to the opacity coupled with or replaced by a more gentle  tachocline flow can be found to 
reproduce the observed sound-speed anomaly illustrated in Fig.~\ref{fig1}.

Before proceeding with a discussion of the results of such an exercise, we recognize that analyses by \citet{asplundetal2004A&A,Asplundetal2005,Asplundetal2009ARA&A}, 
and subsequently \citet{caffauetal3dmodel2008A&A,caffauetal2009A&A,Caffauetal2011SolPh}, have resulted in a
downward revision of $Z_{\rm s}$. 
According to the latest results by \citet{Asplundetal2009ARA&A}
the reduction is approximately 25 per cent, yielding 
$Z_{\rm s}/X_{\rm s} = 0.0181 \pm 0.0008$.%
\footnote{The standard error represents an estimate by Remo Collet
(private communication).}
The effect on the equation of state is small, and has little influence on the
stratification of a calibrated solar model.
It is just the opacity $\kappa$ that is affected. 
\citet{JCD-Houdek2010} computed a solar model with the same physics
as Model S but with the revised value of $Z_{\rm s}$ 
obtained by \citet{Asplundetal2009ARA&A}, and they have
demonstrated explicitly that an intrinsic modification
$\Delta \log \kappa$ to $\kappa$, assumed to depend just on $T$,
can be found that converts its seismic structure to that of Model S.
The opacity modification is otherwise arbitrary, but it is interesting to note
that its value at least near the base of the convection zone 
is comparable with the revision in the iron opacity inferred recently from the
Z-pinch experiment \citep{baileyetalZpinchopacity2015Natur.517...56B,naharpradhan2016PhRvL.116w5003N}.%
\footnote{\citet{DOGLorentz2004AIPC} determined the actual opacity difference beneath
the tachocline between a seismically determined representation of the Sun
and the original Model S.}
The result of a seismic analysis based on this model, Model S\'\, 
in the following, is illustrated in Fig.~\ref{fig1} by the dotted line; 
it is evident that it closely resembles the original Model S.
It appears, therefore, that it behoves us only to model the tachocline anomaly
in its correct location.
The analysis is based on using $\Delta \log \kappa(T)$ as 
obtained by \citet{JCD-Houdek2010} together with the
\citet{Asplundetal2009ARA&A} composition,
using Model S\'\, as a starting point.
We expect that a moderate further slight adjustment to $\kappa$ would then
suffice to remove any additional structural discrepancy at depth that
may ensue. 
We now illustrate the extent to which we have been successful.

\begin{figure}
          \includegraphics[width=9.0cm]{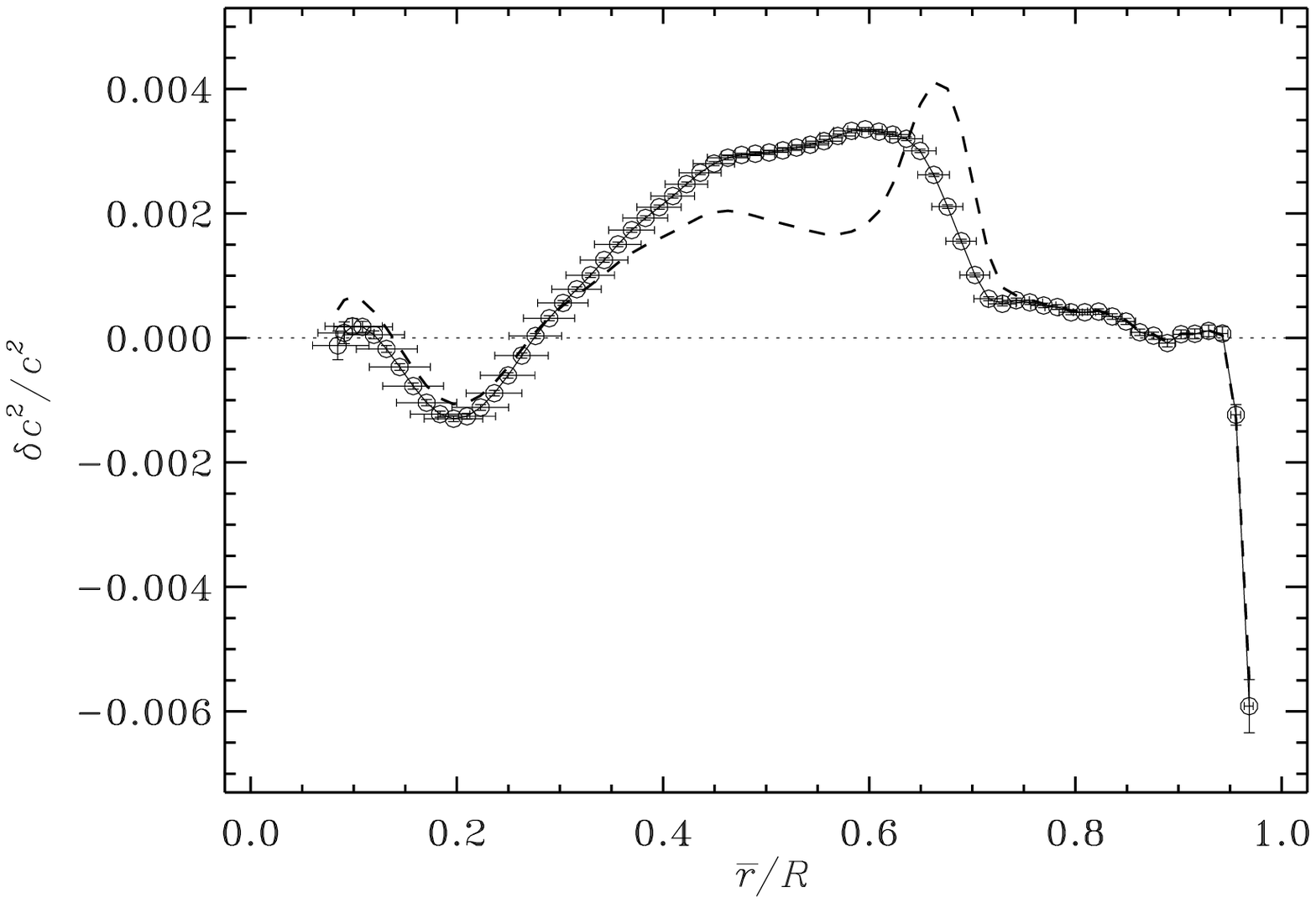}
\caption{
Relative sound-speed difference between the Sun and a model computed with 
the same physics as Model S\'\,, using the \citet{Asplundetal2009ARA&A}
composition and the opacity correction of \citet{JCD-Houdek2010}, 
except that there has been added material
diffusion beneath the convection zone with a diffusion coefficient $D$
according to equation (\ref{eq:1}), with
$D_0 = 150 \, {\rm cm^2 \, s^{-1}}$,
$v_0 = 0.15 \, {\rm g^{-1} \, cm^3}$ and $\alpha = 4.25$.
The surface composition is characterized by $Z_{\rm s}/X_{\rm s} = 0.0181$.
There remains a large-scale discrepancy in the radiative interior which can
be largely removed by a suitable modification to the opacity.
The dashed curve reproduces the deviation from Model S\'\, illustrated 
in Fig.~\ref{fig1} by the dotted curve.
}\label{fig6}
\end{figure}

In the light of our discussions in Sections 3 and 4, we propose
a putative degree of weak material mixing beneath the convectively unstably stratified 
zone to produce a layer in which the chemical 
composition is not completely homogenized.  
That layer is presumed to be in radiative equilibrium. 
To achieve that state we adopt a diffusive process with a diffusion coefficient $D$ given by a generalization of that considered by 
\citet{jcdmpdimdiffusion2007EAS..26..3C}, namely
\begin{equation}
D=D_0\left(\frac{v-v_0}{v_{\rm c}-v_0}\right)^\alpha , \qquad v > v_0 ,
\label{eq:1}
\end{equation}
and $D = 0$, otherwise,
where $v=1/\rho$ and $v_{\rm c}=1/\rho_{\rm c}$, in which $\rho_{\rm c}$
is the density at the base of the convection zone 
($\rho_{\rm c}= 0.1902\, {\rm g \,cm^{-3}}$ in Model S\'\,);
$D_0$, $v_0$ and $\alpha$ are adjustable parameters.  Christensen-Dalsgaard and Di Mauro considered diffusion of chemical species with $v_0=0$ and $\alpha=3$, as had been suggested by \citet{proffittmichaud1991ApJ...380..238P}.    They found that a suitably adjusted value of $D_0$ can go a long way towards removing the acoustic anomaly.  However, in common with a slow tachocline of the form described in Section~3, 
there remained a small sharp dip in $\delta c^2$ immediately beneath the model convection zone.   One might imagine removing that defect  with 
an augmentation of temperature at the base of the convection zone,
produced by either an increase in opacity in the radiative diffusion layer,
or by postulating a shallow layer of rapid entropy mixing.

\begin{figure}
          \includegraphics[width=9.0cm]{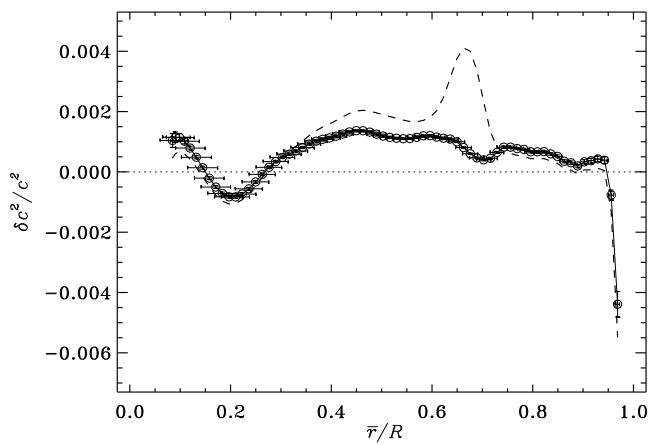}
\caption{
As Fig.~\ref{fig6}, except that
the surface composition is characterized by $Z_{\rm s}/X_{\rm s} = 0.0189$.
The small remaining large-scale discrepancy in the radiative interior can
be largely removed by a suitable modification to the opacity.
}\label{fig7}
\end{figure}

\begin{figure}
          \includegraphics[width=9.0cm]{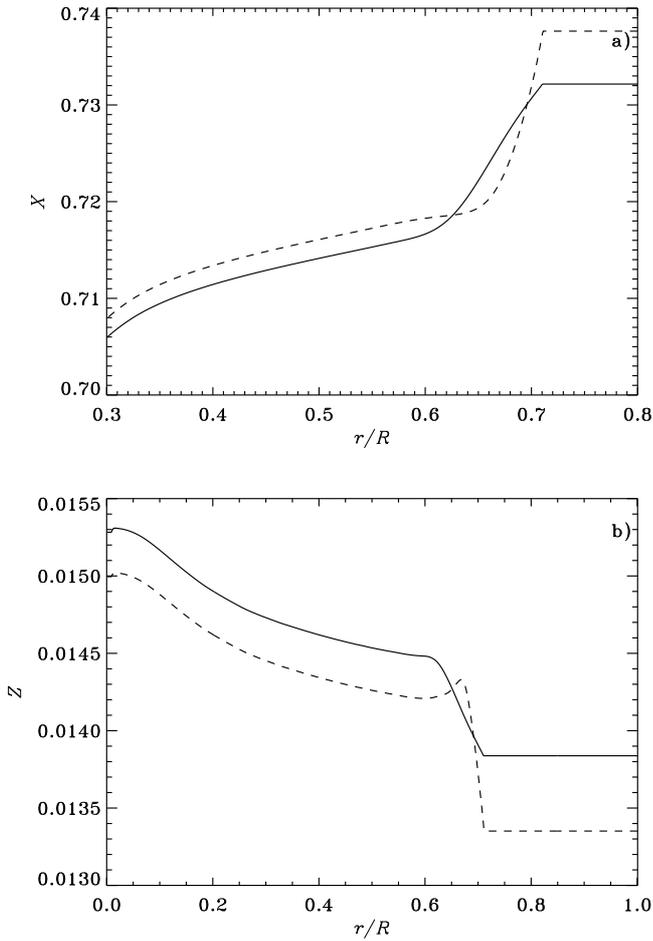}
\caption{
Abundance profiles of hydrogen, $X$, and heavy elements, $Z$, in the model illustrated in
Fig.~\ref{fig7} with 
diffusive mixing,
{\rd maintaining the radiative thermal stratification,}
immediately beneath the convection zone (continuous curves) and in Model S\'\, (dashed curves).
}\label{fig8}
\end{figure}

\begin{figure}
          \includegraphics[width=9.0cm]{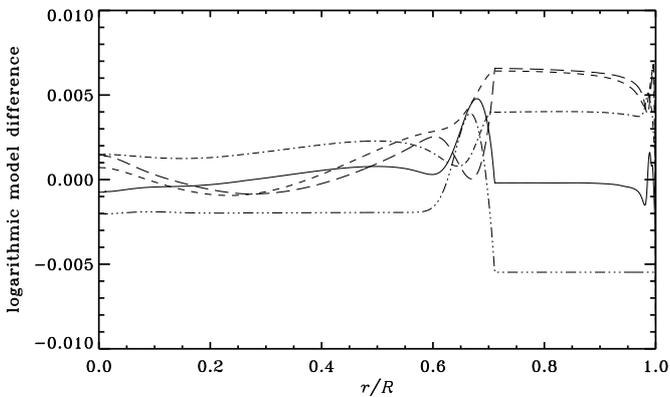}
\caption{
	Differences in ${\rm ln}\, c^2$ (continuous curve), ${\rm ln}\, p$ (short dashed), ${\rm ln} \,\rho$ (long dashed), ${\rm ln} \,T$ (dot-dashed) and $X$ (triple-dot-dashed) between the model illustrated in Fig.~\ref{fig7}
and Model S\'\,.
}\label{fig9}
\end{figure}

Here we simply adopt the diffusion process, and adjust $D_0$, $v_0$ and
$\alpha$ in an attempt to obtain a satisfactory stratification beneath
the convection zone.
We illustrate the outcome in Fig.~\ref{fig6}.
There is no unique way of defining what one might consider to be a `best' fit
to the seismically determined structure.
Here we select a model with $D_0 = 150 \, {\rm cm^2 \, s^{-1}}$,
$v_0 = 0.15 \, {\rm g^{-1} \, cm^3}$ and $\alpha = 4.25$ throughout the main-sequence evolution,
calibrated to have $Z_{\rm s}/X_{\rm s} = 0.0181$ at the present age of the Sun.
This has successfully suppressed the tachocline anomaly,
but at the expense of a somewhat larger deviation in the 
deeper seismically accessible regions of the {\rd radiative interior}.
As argued above, this smooth variation could in principle be suppressed by an
appropriate modification to the opacity.
Here we have all but removed it in a new model obtained by simply augmenting   
$Z_{\rm s}/X_{\rm s}$ in the model of the present Sun to 0.0189,
at the limit of the assumed standard error in the composition determination by
\citet{Asplundetal2009ARA&A}.
The outcome is illustrated in Figs~\ref{fig7} -- \ref{fig9}. 
The tachocline anomaly has almost disappeared.

In Fig.~\ref{fig8} we compare the smoothed 
abundances $X$ and $Z$ of hydrogen and heavy elements in the new model 
with the 
corresponding abundances in Model S\'\,. 
Also, Fig.~\ref{fig9} shows relative differences between the two models.
The overall effect of diffusive mixing is similar to that of the slow and fast tachocline  
discussed in Sections 3 and 4.  The immediate consequence of mixing some of the settling material back into the convection zone, required for reducing the tachocline anomaly, is to reduce $X$ in the convection zone, 
with a further reduction resulting from calibrating the model to a
higher $Z_{\rm s}/X_{\rm s}$; 
this is coupled 
with a concomitant increase in temperature.
The inclusion of diffusive mixing just beneath the convection zone has 
decreased the settling of both hydrogen and heavy elements.
The more gradual decrease of the hydrogen abundance below the convection
zone dominates the sound-speed difference, resulting in a peak which,
when smoothed by the averaging kernels, closely matches the original
tachocline anomaly found in the inversion using Model S\'\,.
The change in calibration also leads to a general increase in $Z$ and hence
opacity, which is counteracted by an increase in temperature,
maintaining the required radiative energy transport.
The increase in temperature also largely balances the increase in
mean molecular mass and hence, unlike the situation in Fig.~\ref{fig6},
leads to a modest change in sound speed in the bulk of the radiative interior,
where also Model S\'\, provides a reasonable match to the helioseismologically
inferred stratification.

\section{Discussion and conclusions}

We have shown that material diffusion immediately beneath the convection zone,
with a suitably chosen diffusion coefficient, can all but remove the
acoustic glitch in solar Model S which arises principally from 
excessive gravitational settling of helium in the model.
The partial chemical homogenization apparent in the spherically averaged
structure to which the seismological analysis is sensitive could be
the result of the tachocline circulation and possible convective 
overshooting, the latter having been caused in part by a true small-scale
mixing and in part by a laminar undulation of the interface between the
fully mixed convection zone and the relatively quiescent radiative interior,
the details of which we have not studied.

We do not claim that we have necessarily determined the cause of the difference 
between Model S and the Sun in the tachocline region.  There are certainly other possibilities, such as 
an error in the procedure adopted to model gravitational settling in Model S,  
an asphericity in the location of the base of the convection zone, or Stokes drift 
and Taylor dispersion by gravity waves generated at the base of the convection zone 
\citep[e.g.,][]{taylor-like_dispersion1992ApJ...401..196K},
none of which have been investigated in detail.

A noteworthy feature in the difference between the solar and model structures, 
which persists in the model illustrated in Fig.~\ref{fig7}, is the systematic 
variation below $r \approx 0.3\,R$.
This is a region where, as illustrated by the horizontal bars,
the resolution of the inversion is comparatively poor; 
in addition, there is substantial correlation between the 
inferences at different values of $\overline{r}$  
\citep[e.g.,][]{rabellosoaresetalinversionparameters1999MNRAS.309.35R},
rendering the apparently smooth variation somewhat misleading.
Even so, it is a common feature of helioseismological inferences, and hence
likely to be of some physical significance.
It is possible that it could be eliminated by a suitable modification
to the opacity.
However, it is interesting that it occurs in a region where the
hydrogen abundance has been substantially modified by nuclear fusion,
raising the possibility that additional weak mixing has occurred 
\citep[e.g.,][]{DOGAGK1990ASSL..159..327G}.

Our goal in the present paper has been to elucidate the possible features
that may affect the tachocline anomaly without introducing other discrepancies
between the inferred solar sound-speed stratification and the model.
This is the background for using data and model physics that have been
extensively studied in the past but are no longer up to date.
A more definitive analysis should be based on the most recent 
observational frequencies, such as those provided by 
\citet{reiter_etal_2015ApJ...803...92R}.
We note, however, that the inverse analysis presented in that paper
shows differences between the Sun and Model S very similar to our starting
point in Fig.~\ref{fig1}.
Furthermore, the model physics, and the assumed surface composition,
should be updated \citep[e.g.,][]{vinyoles_etal_2017ApJ...835..202V}.
Additional constraints on the properties of the tachocline region can
be obtained by carrying out inverse analyses in terms of different
pairs of variables characterizing the model structure;
an interesting analysis in terms of the Ledoux discriminant 
has been presented by \citet{buldgen_etal_2017MNRAS.472L..70B}.
It seems likely that the results of these analyses will still,
as in our preliminary work, point to a combination of modifications
to the opacity and suitable transport processes changing the local
composition \citep[see also][]{song_etal_2017arXiv171002147S}.
On this basis one may then attempt to identify the physical features
requiring improvement in, respectively,
the relevant atomic physics and the treatment of the stellar internal dynamics.

\section*{Acknowledgements}

We thank Maria Pia Di Mauro for making available her code for structure
inversion, {\rd and the referee for helpful comments which have 
contributed to improving the text.}
Funding for the Stellar Astrophysics Centre is provided by
The Danish National Research Foundation (Grant DNRF106). 
The research was supported by the ASTERISK project 
(ASTERoseismic Investigations with SONG and Kepler) 
funded by the European Research Council (Grant agreement no.: 267864).

\bibliography{jcdupdated_references}

\appendix
\section{The sound speed in the convection zone}

The sound speed throughout most of the adiabatically stratified convection zone 
beneath the He ionization zones, where $\gamma_1\simeq 5/3$, is essentially invariant amongst solar models. This is easily appreciated by recognizing that hydrostatic support implies \citep{DOGCatania1984MmSAI}
\begin{equation}
\frac{r^2}{Gm}\frac{{\rm d}\,c^2}{{\rm d}\,r} =\Theta, \label{A1}
\end{equation}
where $\Theta \simeq 1-\gamma_1 \simeq -2/3$  and $m$ is the mass 
enclosed in the sphere of radius $r$,
which is approximately the mass $M$ of the entire star.
Setting $m=M$ permits integration to yield  
\begin{equation}
c^2 \simeq -GM\Theta(r^{-1}-R_{\rm s}^{-1}) \; 
\label{A2} 
\end{equation}
\citep[e.g.,][]{JCD_1986ASIC..169...23C},
where $R_{\rm s}$ approximates the seismic radius of the star, whose height above the photosphere amongst the calibrated models discussed here varies by much less than a typical scale height in the superadiabatic boundary layer, no more than a few 
hundredths per cent of $R_{\rm s}$ itself.  Therefore $c(r)$ is quite well defined in terms of the mass and radius of the Sun.



\end{document}